\documentclass[prl,english,preprintnumbers,amsmath,amssymb,nofootinbib,twocolumn,superscriptaddress]{revtex4-1}

\pdfoutput=1

\usepackage[latin1]{inputenc}
\usepackage{graphicx}
\usepackage{bbm}
\usepackage{amssymb}
\usepackage{amsmath}

\usepackage{dsfont}

\def\0#1#2{\frac{#1}{#2}}

\def\s0#1#2{\mbox{\small{$ \frac{#1}{#2} $}}}

\newcommand{\Tr}{\mathrm{Tr}}

\newcommand{\E}{\mathrm{e}}
\newcommand{\I}{\mathrm{i}}
\newcommand{\be}{\begin{eqnarray}}
\newcommand{\ee}{\end{eqnarray}}

\newcommand{\nn}{\nonumber }

\newcommand{\zgc}{\mathcal Z}

\usepackage{babel}
\makeatother
\begin{document}

\title{Fermi gases with imaginary mass imbalance\\ and the sign problem in Monte Carlo calculations}

\author{Dietrich Roscher}
\affiliation{Institut f\"ur Kernphysik (Theoriezentrum), Technische Universit\"at Darmstadt, 
D-64289 Darmstadt, Germany}
\author{Jens Braun} 
\affiliation{Institut f\"ur Kernphysik (Theoriezentrum), Technische Universit\"at Darmstadt, 
D-64289 Darmstadt, Germany}
\affiliation{ExtreMe Matter Institute EMMI, GSI, Planckstra{\ss}e 1, D-64291 Darmstadt, Germany}
\author{Jiunn-Wei Chen} 
\affiliation{Department of Physics, National Center for Theoretical Sciences and Leung Center for Cosmology and Particle Astrophysics, National Taiwan University, Taipei 10617, Taiwan}
\author{Joaqu\'{\i}n E. Drut} 
\affiliation{Department of Physics and Astronomy, University of North Carolina, Chapel Hill, NC 27599, USA}

\begin{abstract}
Fermi gases in strongly coupled regimes, such as the unitary limit, are inherently challenging for many-body methods.
Although much progress has been made with purely analytic methods, quantitative results require {\it ab initio} numerical 
approaches, such as Monte Carlo (MC) calculations.
However, mass-imbalanced and spin-imbalanced gases are not accessible to MC calculations due to 
the infamous sign problem.
It was recently pointed out that the sign problem, for finite spin imbalance, can be circumvented
by resorting to imaginary polarizations and analytic continuation. Large parts of the phase diagram spanned by temperature and polarization
then become accessible to MC calculations. We propose to apply a similar strategy 
to the mass-imbalanced case, which opens up the possibility to study the associated phase diagram with
MC calculations. In particular, our analysis suggests that a detection of a (tri-)critical point in this phase
diagram is possible. We also discuss calculations in the zero-temperature limit with our approach.
\end{abstract}

\maketitle

{\it Introduction -} Ultracold Fermi gases have attracted considerable interest in the past 15 years. The high level of activity in this field
can be traced back to the fact that quantum many-body phenomena~\cite{RevTheory,RevExp}, ranging from Bose-Einstein condensation 
(BEC) to Bardeen-Cooper-Schrieffer (BCS) superfluidity, have become accessible to experiments in a progressively cleaner and
more versatile way. The (experimental) control parameters are 
the density~$n$ and the s-wave scattering length~$a_{\rm s}$. The latter can be tuned by an external magnetic field
in the presence of a Feshbach resonance. For a sufficiently dilute Fermi gas, the range of the interaction 
effectively represents the smallest length scale in this many-body problem and therefore the dynamics
is completely controlled by the dimensionless parameter~$k^{}_{\rm F} a^{}_{\rm s}$, where the Fermi momentum~$k_{\rm F}$
is determined by the density~$n$.

In the limit of large $s$-wave scattering length, the so-called unitary regime, the only scale left in the problem is the density $n$. 
For these systems, a small expansion
parameter has not yet been identified and most likely does not exist. Therefore, the use of non-perturbative methods
is unavoidable, which makes this limit a major challenge for theoretical approaches~\cite{ZwergerBook}.
On the other hand, experimental studies have achieved high precision in some cases~\cite{PrecisionExp}. Apart from their phenomenological
relevance, these experiments can therefore be used to benchmark theoretical methods. 

In the present work, we restrict ourselves to the case of mass-imbalanced Fermi gases in the unitary regime, i.e. we are
interested in Fermi gases consisting of two species with unequal masses interacting resonantly with each other. 
To study this special class of Fermi gas, great efforts have been made in recent years, both on the experimental~\cite{miexp}
and the theoretical side (see, e. g., Ref.~\cite{revstoof} for a review). Nevertheless, mass-imbalanced Fermi gases remain out of reach
for {\it ab initio} Monte Carlo (MC) calculations due to the infamous sign problem (see, e.~g., Ref.~\cite{Drut:2012md}).
On the other hand, the mass-balanced unitary Fermi gas is free of these complications~\cite{Chen:2003vy,Bulgac:2005pj} 
and shows good agreement with experimental data~\cite{Drut:2011tf}. 
The situation resembles that of spin-imbalanced Fermi gases, which 
are also affected by the sign problem in MC approaches. For the latter, however, an approach
to circumvent the sign problem has been put forward in Ref.~\cite{Braun:2012ww}. 
In analogy to lattice studies of the QCD phase diagram~\cite{latticeQCD}, this strategy employs an analytic continuation
from real to imaginary polarization which opens up the possibility to study large parts of the spin-imbalanced 
finite-temperature phase diagram, potentially including the determination of the location of the (tri)critical point.
 
Inspired by the analytic continuation of the theory in case of spin-imbalanced gases~\cite{Braun:2012ww}, we show here that a similar analytic 
continuation could be used to study mass-imbalanced Fermi gases with {\it ab initio} MC calculations without any kind of
sign problem. In fact, we will demonstrate that, in contrast to the spin-imbalanced case, not only the finite-temperature
phase diagram is accessible in the present case but also the zero-temperature limit. This opens up
the possibility for a detailed comparison of experimental data and results from {\it ab initio} MC calculations, which will help
us enhance our understanding of the dynamics underlying pairing in strongly interacting Fermi gases. {Note that this
strategy of analytically continuing the theory works in any dimension as well as away from unitarity.}

{\it Formalism -} In the following we apply a (standard) mean-field approach to a unitary mass-imbalanced Fermi gas in three (spatial) dimensions,
as also discussed elsewhere (see, e.g., Refs.~\cite{revchevy,revstoof,BGS}). In contrast to earlier works, however, we employ real- and 
imaginary mass imbalances for our study of the phase structure. In particular, we discuss how the phase diagram for imaginary-valued
mass imbalances can be used to determine the ground-state properties of the Fermi gas in the physical case, i.e. for
real-valued mass imbalances. Of course, we do not expect our mean-field approach to yield the exact values
for physical observables. Nevertheless, it can be viewed as the lowest-order approximation and,  
as is the case for the full evaluation of the associated path integral,
it only depends on a single input parameter (e.g. $k_{\rm F}$). This is important because it implies that our results
do not suffer from a parameter ambiguity, but only from uncertainties associated with the mean-field approximation,
(see, e.g., Refs.~\cite{Diehl:2009ma} for a more general discussion of this issue).
Therefore, we expect that our mean-field approach is suitable
to present our approach based on imaginary-valued mass imbalances, and allows us to gain essential insights into the
(analytic) structure of the theory. The latter can then be used to guide future {\it ab initio} MC calculations.

Before discussing our mean-field study of the unitary Fermi gas with imaginary-valued mass imbalance, we
give a more general introduction to the problem. The partition function of a Fermi gas reads
\be
\zgc (T,m_{\uparrow},m_{\downarrow},\bar{\mu},h)=\Tr\left[ \E^{-\beta(\hat{H}-\bar{\mu}(\hat{N}^{}_{\uparrow}+\hat{N}^{}_{\downarrow})
-h (\hat{N}^{}_{\uparrow}-\hat{N}^{}_{\downarrow}))}\right]\,, \label{eq:Z} \nn
\ee
where $\beta=1/T$ is the inverse temperature. We assume that the Hamiltonian~$\hat{H}$ describes 
the dynamics of a theory with two fermion species, denoted by \mbox{$\uparrow$ and $\downarrow$}, interacting only
via a zero-range two-body interaction:
\be
\hat{H}= 
\int d^{3}x \left (
 \sum_{\sigma=\uparrow,\downarrow}
  \hat{\psi}_{\sigma}^{\dagger}(\mathbf{x})\frac{\vec{\nabla}^2}{2m_{\sigma}}\hat{\psi}_{\sigma} (\mathbf{x})
  + \bar{g} \hat{\rho}_\uparrow (\mathbf{x}) \hat{\rho}_\downarrow (\mathbf{x}) \right )\,,\nn
\ee
The operators~$\hat{\rho}^{}_{\uparrow,\downarrow}$ are the particle density operators associated with the two 
fermion species, and $\hat{N}^{}_{\uparrow,\downarrow}$ are the corresponding particle-number operators. 
The masses of the two species are given by~$m_{\uparrow}$ and~$m_{\downarrow}$, respectively. 
For the sake of generality, we have also introduced the average chemical potential~$\bar{\mu}=(\mu^{}_{\uparrow}+\mu^{}_{\downarrow})/2$ 
and the asymmetry parameter~$h=(\mu^{}_{\uparrow}-\mu^{}_{\downarrow})/2$. From now on, however, we consider
the case~$h=0$ and allow only for a finite mass imbalance. To this end, it is convenient to introduce quantities that allow
us to measure the mass imbalance in simple terms:
\be
m_{+}=\frac{4 m_{\uparrow}m_{\downarrow}}{m_{\uparrow}+m_{\downarrow}}\,,\quad
m_{-}=\frac{4 m_{\uparrow}m_{\downarrow}}{m_{\downarrow}-m_{\uparrow}}\,,\quad
\bar{m}=\frac{m_{+}}{m_{-}}\,.
\ee
The dimensionless imbalance parameter~$\bar{m}=(m_{\downarrow}-m_{\uparrow})/(m_{\downarrow}+m_{\uparrow})$ 
measures the relative strength of the mass imbalance. {Note that~$0\leq |\bar{m}| \leq 1$.}

As already mentioned above, MC calculations for unitary fermions with~$h=0$ and~$\bar{m}=0$
can be performed without a sign problem (see e.g. Refs.~\cite{Chen:2003vy,Bulgac:2005pj,Drut:2012md}).
For the case~$h>0$ and~$\bar{m}=0$, it has been pointed out in Ref.~\cite{Braun:2012ww} that this sign
problem can be circumvented by considering an imaginary-valued asymmetry parameter~$h$. This corresponds
to studying the theory with complex-valued chemical potentials. The results from such an MC study can then be
analytically continued to obtain the results for the physically interesting case of a real-valued asymmetry parameter.
For the case~$h=0$ and~$\bar{m}>0$, it turns out that a similar approach can be used to avoid the appearance
of the sign problem in MC calculations. Assuming that~$\bar{m}_{-}$ is imaginary-valued, it is indeed straightforward to show that the
fermion determinants appearing in the probability measure are complex conjugates of one another, provided that 
the parameter~$m_{+}$ is still considered to be real-valued and~$m_{\uparrow}^{\ast}=m_{\downarrow}$. For convenience, we
define~$\bar{m}=\I \bar{m}^{}_{\rm I}$ with $\bar{m}^{}_{\rm I}\in \mathbb{R}$. Fermi gases with imaginary mass imbalance can thus be studied using standard MC techniques. As in the case of an imaginary-valued spin-imbalance parameter, the actual
results of physical interest are to be found by 
analytically continuing the partition function to real-valued mass imbalance, to obtain the original partition function
$\zgc (T,m_{\uparrow},m_{\downarrow},\bar{\mu},h)$ with~$m_{\uparrow,\downarrow}\in \mathbb{R}$. From the latter we can, in principle, 
extract all experimentally observable equilibrium quantities.
As we discuss below in more detail, the zero-temperature limit is also accessible in the present case, {which
is not possible in the imaginary $h$ case.}

{\it Mean-field analysis -} To illustrate our imaginary mass imbalance approach and to gain deeper insights into the analytic structure
of the theory (which is required to guide future MC studies), we employ a mean-field study with complex-valued 
masses~$m_{\uparrow,\downarrow}$, such that~$m_{\uparrow}^{\ast}=m_{\downarrow}$. The real and imaginary
parts of these masses can then be tuned to obtain a certain given value for the imbalance parameter~$\bar{m}$.
In order to compute the order-parameter potential for U($1$) symmetry breaking in the mean-field approximation, 
we employ the path-integral representation of~$\zgc$:
\be
\zgc = \int {\mathcal D}\psi^{\dagger}{\mathcal D}\psi\, \E^{-{S} [\psi^{\dagger},\psi]
 }\,,\nn
\ee
where
\be
&& S [\psi^{\dagger},\psi]=  \int d\tau\int d^3 x\, \bigg\{ \psi^{\dagger}
\left( \partial _{\tau} - \frac{1}{m_{+}}\vec{\nabla}^{\,2} -\bar{\mu} \right) \psi \nn\\
&&  \quad - \frac{1}{m_{-}} \left(  \psi^{\ast}_{\uparrow}\vec{\nabla}^{\,2}\psi^{}_{\uparrow} 
-  \psi^{\ast}_{\downarrow}\vec{\nabla}^{\,2}\psi^{}_{\downarrow}\right)
 + \, {\bar{g}}(\psi^{\dagger}\psi) (\psi^{\dagger}\psi)
\bigg\}\,,\label{eq:action}
\ee
and $\psi^{\rm T}=(\psi^{}_{\uparrow},\psi^{}_{\downarrow})$ and~$\bar{g}$ denotes the bare four-fermion coupling. 
The latter is related to the scattering length~$a_{\rm s}$ by
\be
\bar{g}^{-1} = \Lambda g^{-1} =  {\frac{1}{8\pi}\left( a^{-1}_{\rm s}-c_{\rm reg.}\Lambda\right)}.
\, 
\ee
Here,~$\Lambda$ denotes the ultraviolet cutoff and the constant~$c_{\rm reg.}>0$ depends on the 
regularization scheme. We choose units such that~$m_{+}=1$ corresponding to $2m=1$ for $m_{\uparrow}=m_{\downarrow}=m$.
This choice also implies that~\mbox{$\bar{m}=1/m_{-}$}.

To compute the order-parameter potential from the path integral, we introduce an 
auxiliary scalar field~$\varphi \sim  g^{}_{\varphi}\, \psi_{\uparrow}\psi_{\downarrow}$,
where the parameter~$g^{}_{\varphi}$ is chosen to reproduce the four-fermion term in the action. The fermion
fields then only appear bilinearly in the action and can be integrated out. Assuming a homogeneous
ground state, the resulting fermion determinants can be computed straightforwardly, eventually yielding
the order-parameter potential:
\be
\beta U(\varphi)&=&-2\beta\bar{\mu}|\varphi|^2
\!-\! \int\frac{d^3q}{(2\pi)^3}\ln
\bigg[
\cosh\left(\beta\bar{m} {\vec{q}^{\,2}}\right) \nn\\
&&
+  \cosh\left(\beta\sqrt{\left({\vec{q}^{\,2}} - \bar{\mu}\right)^2 + g_\varphi^2|\varphi|^2}\right)\bigg]
\,.
\label{eq:orderpot}
\ee
{For conciseness,}
we have dropped the standard terms that are required to regularize the potential (see, e.g., Refs.~\cite{revchevy,revstoof}). 

The order-parameter potential~$U$ and the grand canonical potential~$\Omega$ are directly related,~$\Omega=VU(\varphi^{}_0)$, where~$V$ 
is the volume of the system and~$\varphi^{}_0$ denotes the value of $\varphi$ that
minimizes the potential. In the ground state, we can identify~$g_{\varphi}^2 |\varphi^{}_0|^2$ with the fermion gap~$\Delta$.
The latter serves as an order parameter for spontaneous U($1$) symmetry breaking associated with a superfluid ground state.
From the grand canonical potential we can extract all thermodynamic observables.
As it should be, our results for dimensionless (universal) quantities in the unitary limit, such as the critical 
temperature~$T_{\rm c}(\bar{m})/\bar{\mu}$ for the superfluid transition or 
the ground-state energy~$E(\bar{m})/\bar{\mu}$, are independent of~$\bar{\mu}$ and~$g_{\varphi}$.

Next, we discuss some of the analytic properties of the theory.
In the case of~$\bar{m}=0$ and $h>0$, it was found that the theory is 
$2\pi$-periodic in $\beta h_{\rm I}$, where~$h=\I h_{\rm I}$ and~$h_{\rm I}\in \mathbb{R}$ (see Ref.~\cite{Braun:2012ww}). 
Using imaginary spin-imbalances, the accessible spin asymmetries are therefore bound to values~$\beta h^{}_\text{I} < \pi$.
For the mass-imbalanced case, we do not have a corresponding simple periodic behavior in~$\bar{m}$,
as the parameters for mass imbalance and spin imbalance are associated with different operators (cf. Eq.~\eqref{eq:action}).
On the contrary, we find that also the zero-temperature limit is accessible in the case of imaginary-valued
mass asymmetries.

\begin{figure}[t]
\includegraphics[width=\columnwidth]{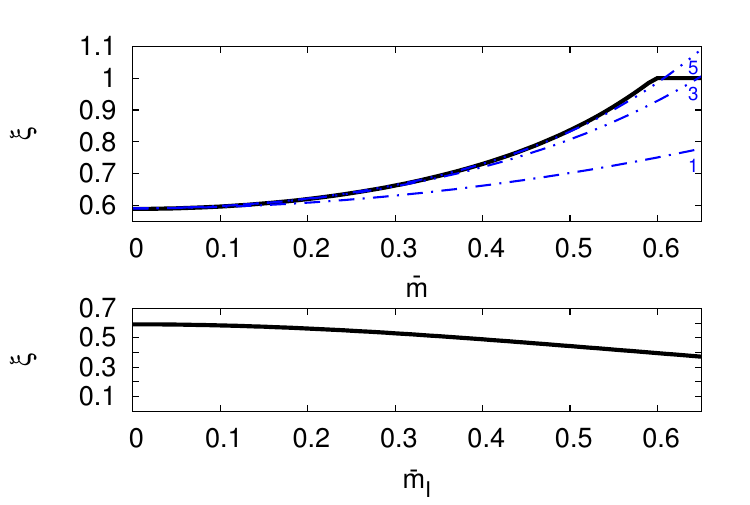}
\caption{\label{fig:1}(color online) 
Upper panel: Bertsch parameter as a function of~$\bar{m}$. The (blue) thin dashed-dotted lines 
are analytic continuations obtained from a Taylor expansion of the Bertsch parameter in~$\bar{m}_{\rm I}^2$ 
up to order~$N_{\rm max}$ (indicated by the numbers, see text).
Lower panel: Bertsch parameter as  a function of~$\bar{m}_{\rm I}$.
}
\end{figure}

The analytic continuation from imaginary-valued to real-valued mass asymmetries trivially requires the theory
to be analytic in a finite domain around the symmetric point $\bar{m}=0$. For given fixed values of~$\varphi$, $T$, and~$\mu$,
we find that the order-parameter potential~\eqref{eq:orderpot} can indeed be expanded in powers of~$\bar{m}$. From the analytic structure
of the integrand in Eq.~\eqref{eq:orderpot}, it follows that a lower bound~$r_{\rm min}$ for the radius of convergence~$r_{\bar{m}}$ 
of such an expansion is given by
\be
r_{\rm min}(|\Phi|^2) = \sqrt{\frac{ \beta^2 |\Phi|^2 +\pi^2 }{\beta^2|\Phi|^2 +\pi^2 + \beta^2\bar{\mu}^2 } }\,,
\ee
where~$|\Phi|^2=g_{\varphi}^2|\varphi|^2$. Notice that this lower bound remains finite as $\beta \to \infty$.
To compute physical observables~$\mathcal O$, such as the density or the ground-state energy, 
we need to evaluate the potential at its (global) minimum~$\varphi_0$.
The radius~$r_{\bar{m},{\mathcal O}}$ for an observable~$\mathcal O$ is then 
{bounded from below by~$r_{\rm min}(\Delta = g_{\varphi}^2|\varphi_0|^2)$,}
provided that the integrand in Eq.~\eqref{eq:orderpot} is holomorphic in~$\Delta$ and~$\Delta$ is holomorphic in~$\bar{m}$. 
For~$\bar{m}<\bar{m}_{\rm c}(T)$, we expect this to be the case. 
Here,~$\bar{m}_{\rm c}(T)$ denotes the critical mass imbalance below which the gap is finite for a given temperature~$T$. 
The actual radius of
convergence may well be larger than this lower bound since the analytic properties of the potential may be improved by the momentum integration
in Eq.~\eqref{eq:orderpot}. {In the superfluid phase, on the other hand, an upper bound}
for the radius of convergence is given by~$\bar{m}_{\rm c}(T)$, i.e. $r_{\bar{m},{\mathcal O}}<\bar{m}_{\rm c}(T)$. For the {\it Bertsch} parameter~$\xi$
at~$T=0$, it indeed appears to be the case that~$r_{\bar{m},\xi}\approx\bar{m}_{\rm c}$ (see Fig.~\ref{fig:1} and our discussion below).

In the high-temperature phase, i.e. $T\gg T_{\rm c}(\bar{m}\!=\!0)$, the gap is zero for~$\bar{m}_{({\rm I})}\in[0,1]$ and we have~$r_{\bar{m},{\mathcal O}} \leq 1$. 
For the two-dimensional Fermi gas, one can indeed show analytically 
that~$r_{\bar{m},{\mathcal O}}=1$ in the high-temperature limit.
In any case, from the point of view of MC calculations, these findings suggest to compute physical observables 
as function of~$\I\bar{m}_{\rm I}$ with standard techniques without suffering from the sign problem. 
The results of physical interest, i.e. for real-valued~$\bar{m}$, can then be
obtained by fitting the MC data to a polynomial ansatz in~$\bar{m}_{\rm I}$ and analytically continuing this polynomial to the real axis. 
Note that only even powers of~$\bar{m}$ can contribute as the partition function is invariant under the transformation~$\bar{m}\to -\bar{m}$.

{\it Analytic continuation -} Let us now demonstrate the analytic continuation from imaginary-valued to real-valued mass imbalances.
For simplicity, we fit the data of a given observable~$\mathcal O$, such as the {\it Bertsch} parameter or the critical temperature,
to the ansatz ${\mathcal O}=\sum_{n=0}^{N_{\rm max}}C^{(n)}_{\mathcal O} \bar{m}^{2n}_\text{I}$, 
where $C^{(n)}_{\mathcal O}$ are constants determined by the fit to the data and $N_{\rm max}$ represents the 
truncation order. The latter is limited by the amount of available data. Here, we have assumed that~$\mathcal O$ 
has been made dimensionless with, e.g., a suitable power of~$\bar{\mu}$. From a simple analytic continuation 
of this polynomial, one then obtains the dependence of~$\mathcal O$ on~$\bar{m}$. Within our analytic study, we can 
easily check the feasibility of such a procedure. In Fig.~\ref{fig:1}, we show our results for the {\it Bertsch} parameter
as a function of~$\bar{m}$
as obtained from such a fit procedure. Here, we have defined the Fermi energy~$\epsilon_{\rm F}$ entering the definition 
of the {\it Bertsch} parameter~$\xi=\bar{\mu}/\epsilon_{\rm F}$ {as follows
\be
\epsilon_{\rm F}\!=\!\left(1\!-\!\bar{m}^2\right)\left[(1\!-\!\bar{m})^{\frac{3}{2}} \!+\!  (1\!+\!\bar{m})^{\frac{3}{2}}
\right]^{-\frac{2}{3}}\!(6\pi^2 n)^{\frac{2}{3}}
\,,
\ee 
where}~$n=n_{\uparrow}+n_{\downarrow}$ is the total density. With this definition of the {\it Bertsch} parameter, we 
have~$\xi = 1$ for the free Fermi gas at~$T=0$ and~$\bar{m}\in [0,1]$. For the interacting gas at~$T=\bar{m}=0$, we recover the 
standard {mean-field result~$\xi =0.59\dots$~\cite{ZwergerBook,revstoof}.}
For the fit presented in Fig.~\ref{fig:1} we have used the results for~$\xi$ for 101 equidistant values of~$\bar{m}_{\rm I}\in [0,1]$. 
For~$N_{\rm max}=5$, we already observe good agreement between the analytically continued polynomial and the
exact (mean-field) result for~$\bar{m} \lesssim \bar{m}_{\rm c}$. We add that a polynomial ansatz
is the simplest choice for such a fit. Of course, more elaborate functions, such as Pad\'{e} approximants, can also be employed.

%

With our imaginary mass-imbalance approach, it is also possible to study the 
thermal properties, as the temperature dependence of the radius of convergence already implies.
The critical temperature~$T_{\rm c}(\bar{m})$ can be computed by studying the gap as a function of the temperature for a given value of~$\bar{m}$.
To be more precise,~$T_{\rm c}(\bar{m})$ is the lowest temperature for which the gap vanishes identically.
Since the critical temperature is defined implicitly by the gap, it appears to be impossible to
derive the radius of convergence for~$T_{\rm c}(\bar{m})$ analytically. In the following, we shall 
assume that the radius of convergence for the critical temperature is finite, as suggested by the analytic properties of the potential. 
Moreover, we expect that the radius of convergence for~$T_{\rm c}(\bar{m})$ is bounded from above by the radius of convergence of the potential.
These statements {are in accordance with our numerical results.}
\begin{figure}[t]
\includegraphics[width=0.95\columnwidth]{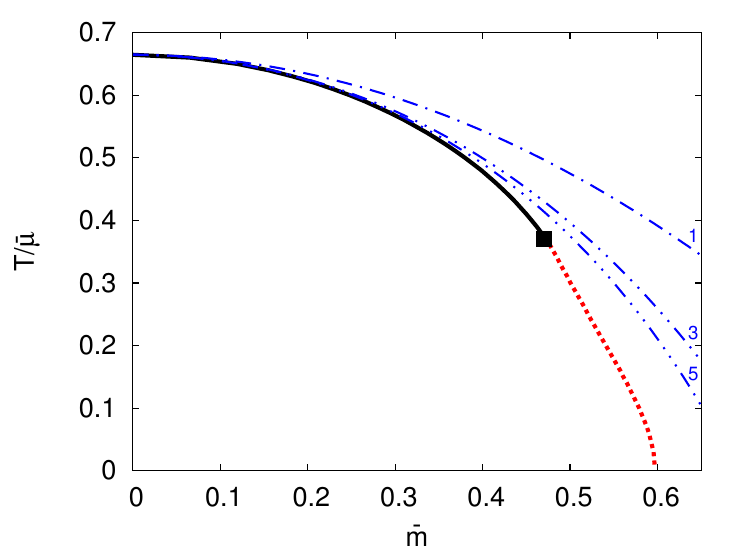}
\caption{\label{fig:2}(color online) 
Phase diagram of an ultracold Fermi gas at unitarity in the ($T$,$\bar{m}$) plane. 
The solid (black) curve is a line of second-order phase transitions, which ends at a (tri)critical point 
$(\bar{m}_{\rm cp},T_{\rm cp}/\bar{\mu})$ marked by the black square. For~$\bar{m}>\bar{m}_{\rm cp}$, we find a line of first-order transitions. 
The (blue) thin dashed-dotted lines 
are analytic continuations obtained from a Taylor expansion 
of the phase boundary in~$\bar{m}_{\rm I}^2$ 
up to order~$N_{\rm max}$ (indicated by the numbers, see text).
}
\end{figure}
In Fig.~\ref{fig:2} we show the mean-field phase
diagram in the $(T,\bar{m})$ plane. 
For~$\bar{m}<\bar{m}_{\rm cp}$, the finite-temperature phase transition is found to be of second order, whereas
it is of first order for~$\bar{m}>\bar{m}_{\rm cp}$. {We stress that
we are not interested in a study of inhomogeneous phases (neither Sarma-type nor FFLO-type) in the present work.} We only discuss the phase boundary between
the U($1$)-symmetric phase and the phase associated with a superfluid (homogeneous) ground-state. The (tri)critical
point~$(T_{\rm cp}/\bar{\mu},\bar{m}_{\rm cp})$ is located within the radius of convergence of the order-parameter potential 
when expanded in powers of~$\bar{m}$. 
In fact, we find that the radius of convergence is about twice as large as the value of the mass-imbalance associated with the (tri)critical point.
The latter observation is intriguing since it suggests that the (tri)critical 
point is potentially within the reach of lattice MC calculations based on an imaginary mass-imbalance. 

\begin{figure}[t]
\includegraphics[width=0.95\columnwidth]{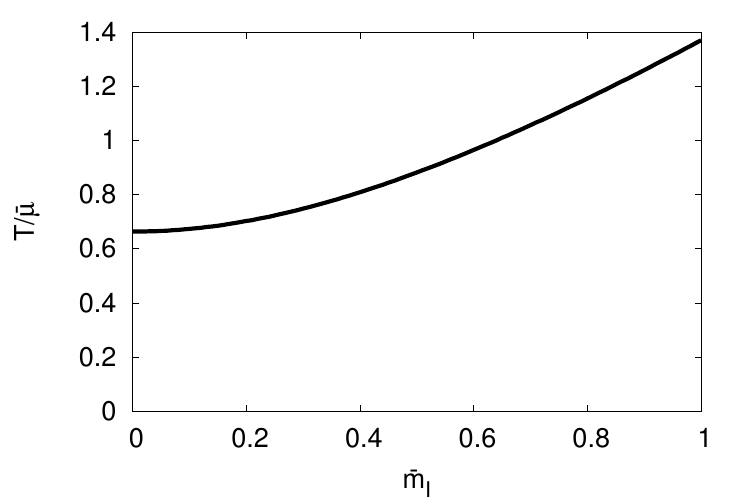}
\caption{\label{fig:3} 
Phase diagram in the ($T$,$\bar{m}^{}_{\rm I}$) plane. 
The solid line is a line of second-order phase transitions below which the fermion gap is finite. 
Note that we do not find a line of first-order phase transitions in this case.
}
\end{figure}

{\it Connection to lattice calculations -} From the point of view of (lattice) MC calculations, one would first compute the phase diagram in the $(T,\bar{m}_{\rm I})$ plane 
with standard techniques without suffering from the sign problem.
In Fig.~\ref{fig:3}, we show the corresponding phase diagram as obtained from our mean-field approximation.
We find that the phase transition line is of second order only and the phase
transition temperature is finite for $\bar{m}_{\rm I}\leq 1$. Thus, the (tri)critical 
point leaves no obvious trace of its existence in this phase diagram. At second glance, however, the information of the
existence of this point is encoded in the shape of the potential at~$(\bar{m}_{\rm cp},T_{\rm cp}/\bar{\mu})$.
In analogy to the case of imaginary spin imbalance discussed in Ref.~\cite{Braun:2012ww} 
as well as to relativistic fermion models~\cite{Karbstein:2006er}, we expect that 
the analytic continuation of the phase transition line can only give the correct behavior up to the (tri)critical point. 
From the analytic continuation of the (full) order-parameter potential, however, the 
physical phase diagram in the whole domain defined by the radius of convergence can be recovered, 
including the line of first-order transitions. 

In Fig.~\ref{fig:2}, we also show the results for the finite-temperature phase boundary in the $(T,\bar{m})$ plane as obtained from
the analytic continuation of the function~$T_{\rm c}(\bar{m}_{\rm I})$ depicted in Fig.~\ref{fig:3}. For the analytic continuation, we have 
employed polynomials in~$\bar{m}^2_{\rm I}$ up to order~$N_{\rm max}$ (see our discussion above).
For the fit to the data, we have used the results for~$T_{\rm c}(\bar{m}_{\rm I})/\bar{\mu}$ for 101 equidistant values of~$\bar{m}_{\rm I}\in [0,1]$.
For~$N_{\rm max}=5$ and~\mbox{$\bar{m}\lesssim 0.4$}, we already observe good agreement between the analytically continued polynomial and the
exact (mean-field) result, see Fig.~\ref{fig:2}. 

{\it Summary -} We have studied the dynamics of mass-imbalanced Fermi gases
using complex-valued fermion masses. We have argued that this approach allows to 
avoid the sign problem in MC calculations and therefore opens up the possibility
for {\it ab initio} MC studies of mass-imbalanced Fermi gases, in particular at unitarity. We have found that the 
(tri)critical point is in principle within reach in this framework. Moreover,
the zero-temperature limit is directly accessible as well, at least up to a certain value of~$\bar{m}$
implicitly determined by the analytic properties of the (full) theory.
As studies of mass-imbalanced Fermi gases are highly challenging on both the experimental
and theoretical side, our present study constitutes an important development of the recent work 
of Ref.~\cite{Braun:2012ww}. Based on our complex-masses approach, future MC studies may now indeed
allow us to gain deep insights into the collective dynamics underlying strongly-interacting Fermi gases which 
is of great importance for a large variety of systems, ranging from  
superconducting materials in condensed-matter physics to the nuclear many-body problem.
 

{{\it Acknowledgements -}} J.B. and D.R. acknowledge support by the DFG under Grant BR 4005/2-1. Moreover, JB acknowledges support
by HIC for FAIR within the LOEWE program of the State of Hesse.
{J.-W.C. is supported by the NSC (99-2112-M-002-010- MY3) of ROC and CASTS \& CTS of NTU.}

\bibliographystyle{h-physrev3}

\end{document}